\documentclass[aps,prl,a4,twocolumn,superscriptaddress,nofootinbib,showpacs,showkeys,preprintnumbers]{revtex4-1}
\usepackage{amsfonts}
\usepackage{mathrsfs}
\usepackage{amssymb}
\usepackage{amsmath}
\usepackage{graphicx,epsfig}
\usepackage{pstricks}
\usepackage{slashed}
\usepackage{dcolumn}%

\def\bs{\begin{small}}
\def\es{\end{small}}
\def\be{\begin{equation}}
\def\ee{\end{equation}}
\def\bea{\begin{eqnarray}}
\def\eea{\end{eqnarray}}
\def\bean{\begin{eqnarray*}}
\def\eean{\end{eqnarray*}}
\def\bary{\begin{array}}
\def\eary{\end{array}}
\def\bit{\begin{itemize}}
\def\eit{\end{itemize}}

\def\su5u1{SU(5) \times U(1)}
\def\fsu5u1{SU(5) \times U(1)'}
\def\so10{SO(10)}
\def\sq20{SO(10) \times SO(10)}

\def\bwt{\begin{widetext}}
\def\ewt{\end{widetext}}
\def\be{\begin{equation}}
\def\ee{\end{equation}}
\def\bea{\begin{eqnarray}}
\def\eea{\end{eqnarray}}
\def\bean{\begin{eqnarray*}}
\def\eean{\end{eqnarray*}}
\def\bary{\begin{array}}
\def\eary{\end{array}}
\def\bit{\begin{itemize}}
\def\eit{\end{itemize}}

\def\su5u1{SU(5) \times U(1)}
\def\fsu5u1{SU(5) \times U(1)'}
\def\so10{SO(10)}
\def\sq20{SO(10) \times SO(10)}

\usepackage{amssymb}

\begin{document}

\setlength{\parskip}{0.1cm}


\title{\large Polarized window for left-right symmetry and a right-handed neutrino 
at the Large Hadron-Electron Collider}

\author{Subhadeep Mondal}
\email{subhadeepmondal@hri.res.in}
\affiliation{Regional Centre for Accelerator-based Particle Physics, Harish-Chandra Research Institute,
Chhatnag Road, Jhusi, Allahabad 211019, India}
\author{Santosh Kumar Rai}
\email{skrai@hri.res.in}
\affiliation{Regional Centre for Accelerator-based Particle Physics, Harish-Chandra Research Institute,
Chhatnag Road, Jhusi, Allahabad 211019, India}

\begin{abstract}
The breaking of parity, a fundamental symmetry between left and right is best understood in the 
framework of left-right symmetric extension of the standard model. We show that the production of a 
heavy right-handed neutrino at the proposed Large Hadron-Electron Collider (LHeC) could 
give us the most simple and direct hint of the scale of this breaking in left-right symmetric theories. 
This production mode gives a lepton number violating signal with $\Delta L=2$ which is very clean 
and has practically no standard model background. We highlight that the right-handed nature of 
$W_R$ exchange which defines the left-right symmetric theories can be confirmed by using a polarized 
electron beam and also enhance the production rates with relatively lower beam energy.  
\end{abstract}



\maketitle



The standard model (SM) picture of particle physics has almost already seen its great success at the Large 
Hadron Collider (LHC) with the discovery of the Higgs boson. Notwithstanding the success, what has been 
more intriguing is the lack of any clear hint of physics beyond the SM (BSM) within the LHC data yet. 
Of course there have been some excitement with reported excesses at the kinematic thresholds of the 
recently concluded run-I of the LHC. But these have hardly been conclusive of 
anything and a better 
picture will definitely emerge at the run-II of LHC. However, we are already aware that the SM picture is 
incomplete and one of the clearest signs of BSM has been the existence of non-zero neutrinos mass. 
Whether the neutrinos are Majorana 
or Dirac type is yet to be determined but one definitely needs to invoke physics beyond the SM to generate 
mass for the neutrinos. The most popular of the lot is the so called {\it seesaw mechanism} \cite{csaw} 
where light neutrino masses result after the heavier degrees have been integrated out. The Majorana 
nature of the neutrino would lead to a $\Delta L=2$ lepton number violating (LNV) process with its clear 
imprint in processes such as the neutrinoless double beta decay ($0\nu\beta\beta$). 
The possibility of observing such a process at collider experiments such as the LHC is quite difficult and 
requires the seesaw scale to be quite low at around the TeV scale and a heavy 
Majorana neutrino at less than a few hundred GeV \cite{Han:2006ip}. The production of the right-handed 
neutrinos however get enhanced at LHC with new gauge bosons in the theory \cite{Ferrari:2000sp,Huitu:2008gf,Dev:2013wba}. Left-Right (LR) models 
\cite{Pati:1975, Mohapatra:1974, Senjanovic:1975, Senjanovic:1978, Mohapatra:1979, Mohapatra:1980} 
based on the gauge group  
\begin{equation}
{\mathcal G}_{LR} = SU(3)_C \otimes SU(2)_L \otimes SU(2)_R \otimes U(1)_{B-L}.
\end{equation}
were the first which naturally led to the existence of a right-handed neutrino because of the symmetry
and explain the existence of tiny neutrino mass. They also lead to a clear understanding of why parity 
is maximally violated in weak interactions and not introduced in an \emph{ad hoc} manner in the theory.  
Thus a knowledge of the scale of this breaking would help understand one of the fundamental questions
within the SM picture. In this Letter we focus on the minimal version of the left-right symmetric theory. 
Although unrelated, it is quite interesting to note and worth pointing out that most of the new ideas that 
seem to be addressing the reported excesses at run-I of 
LHC  \cite{Khachatryan:2014dka,Aad:2015owa} have taken refuge in 
the framework of left-right symmetry \cite{Deppisch:2014qpa,Brehmer:2015cia,Dev:2015pga}. 
This study underlines the prospects of uncovering such a specific symmetry at the proposed LHeC. 

LHeC will be the second $e^-p$ collider after HERA supposed to be built at the LHC 
tunnel \cite{AbelleiraFernandez:2012cc, Bruening:2013bga}. The design is planned so as to collide an 
electron beam with a typical energy range, 60-150 GeV with a 7 TeV proton beam producing a 
parton level center-of-mass energy close to $\sqrt{\hat{s}} \simeq 1.3$ TeV. It is expected to achieve 
an 100 $fb^{-1}$ integrated luminosity per year. One particular aspect of LHeC that can be very crucial is 
the availability of a polarized electron beam, which we show, can help in probing the left-right symmetry in nature directly. 
 
\emph{Left-right symmetry}: 
The left-right (LR) symmetry model keeps the SM fermion content the same but extends the SM gauge 
symmetry by another $SU(2)$. The minimal model is based on the gauge group ${\mathcal G}_{LR}$.
The quarks and leptons are completely LR symmetric with the following representations under 
${\mathcal G}_{LR}$:
\begin{eqnarray*}
Q_{L} \sim (3,2,1,1/3),  \hspace{0.2in}     \ell_{L} \sim  (1,2,1,-1) \\
Q_{R} \sim (3,1,2,1/3), \hspace{0.2in}     \ell_{R} \sim  (1,1,2,-1)
\end{eqnarray*}
The formula for the electromagnetic charge ($Q$) is given by
\begin{equation}
Q = T_{3 L} + T_{3 R} + {B - L \over 2}\,
\label{qlr}
\end{equation}
where $T_{3L/3R}$ are the Isospin generators of $SU(2)_{L/R}$.
The symmetry breaking pattern of the LR gauge symmetry down to the SM gauge group and finally to 
$U(1)_{em}$ is brought about by giving VEV to the
the $SU(2)_{L,R}$ triplets $\Delta_L (3,1,2)$ and $\Delta_R (1,3,2)$ and a Higgs bi-doublet $\Phi (2,2,0)$ 
where the charges in parentheses are according to the $SU(2)_L \times SU(2)_R \times U(1)_{B-L}$ 
quantum numbers. 

The VEV structure responsible for the above symmetry breaking is as follows:
\begin{equation}
\langle \Delta_L\rangle = 0 \; , \; \langle \Delta_R\rangle = \left[ \begin{array}{cc} 0 & 0 \\
v_R & 0 \end{array}\right] \; , \;  \langle \Phi\rangle = \left[ \begin{array}{cc} v_1 & 0 \\
0 & v_2\, {\rm e}^{i\alpha}\end{array}\right]
\label{vev}
\end{equation}
where $v_{1,2}, v_R$ are real and positive with $v_R \gg v_{1,2}$. Note that the 
LR symmetry gets broken down to SM with $\langle \Delta_R\rangle $ while the SM 
symmetry breaking happens when the bi-doublet acquires a VEV  $\langle \Phi\rangle$.
The $W$ boson has a mass given by 
$M_W^2 = g_L^2 v^2 \equiv g_L^2 (v_1^2 + v_2^2)$ where $g_L$ denotes the $SU(2)_L$ gauge
coupling. The $\Delta_L$ develops a tiny VEV $\langle\Delta_L\rangle\propto v^2/v_R$ which is 
crucial for the light neutrino masses.

The heavy right handed gauge boson masses (neglecting mixing) are given by
\begin{eqnarray}
& M_{W_R}^2 \simeq   g^2 \, v_R^2  , \,\,\,\,\,\, 
M_{Z_R}^2 \simeq    2 ( g^2 + g_{B-L}^2 ) \, v_R^2 
\end{eqnarray}
where the gauge couplings of hypercharge ($g_Y$) and
$g_{B-L}$ are related through $1/g_Y^{2}=1/g_R^{2}+1/g_{B-L}^{2}$.  
Note that due to LR symmetry and Parity (P) being a good symmetry, we have set $g_L=g_R=g$. 
Under this assumption the $M_{Z_R}\simeq 1.7 M_{W_R}$ and therefore the 
$W_R$ discovery would give us a clear and more importantly the first hint of the LR symmetry breaking scale. With the scale of LR breaking around a few TeV one can generate large Majorana masses for the
right-handed neutrinos. This leads to the seesaw mechanism and generate tiny non-zero neutrino masses
for the active neutrinos.   
For more details on the spectrum including that of right-handed neutrinos, and discussion on the limits of the LR scale, we refer the readers to Ref.\cite {Maiezza:2010ic,Chen:2013fna, Roitgrund:2014zka}.  
In fact for our case, in the minimal LR symmetry model, the heavy neutrinos masses are with similar 
hierarchy as the light ones ($M_N \propto m_\nu$).  

\emph{LR symmetry at LHC}: Irrespective of the theoretical constraints on the $M_{W_R}$ of 
about 2.5 TeV \cite{Zhang:2007da, Beall:1981ze},  the LHC data is set to improve upon the 
direct search limits of $M_{W_R} \gtrsim 3 \text{ TeV}$ \cite{Khachatryan:2014dka} at 95\% C.L. obtained at 
LHC with $\sqrt{s}=8$ TeV. With improved center-of-mass energy and higher luminosity, this reach will only 
improve further with a possibility of discovery for a sub 5-6 TeV $W_R$ \cite{Ferrari:2000sp, Maiezza:2010ic}. 
In addition to that it is also expected to exclude a large mass region for the heavy 
right-handed neutrino, as $W_R \to N_R \ell$ gives a dilepton + dijet channel which can be used to 
reconstruct the $W_R$ and $N_R$ mass using $M_{\ell\ell jj}$ and $M_{\ell jj}$ invariant mass distributions 
respectively.
However one cannot rule out the possibility that a sequential $W'$ with couplings to 
left-handed fermions only, might also lead to such a signal. A clear argument in favor of the LR symmetry and 
the signal arising from the production of a $W_R$ boson will be if one can verify the nature of the coupling 
of $W_R$ with the SM fermions.
This is the prime objective that we wish to address through this Letter and 
present a study of the simplest process allowed where the LR scale is directly involved, at the proposed 
LHeC. To do this, we only require to highlight the charged-current interaction Lagrangian in the LR model. 
This has both left and right-handed charged gauge bosons with their corresponding fermionic interactions 
in the mass eigenstate basis:
\begin{align}
\mathcal L_{W} & = & \frac{g}{\sqrt{2}} \left(
\bar f_L V_{L}^\dag  \slashed{W}\!_L f'_L +
\bar f_R  V_{R}^\dag \slashed{W}\!_R f'_R\right)
+\text{h.c.}\, 
\label{eq:lagrangian}
\end{align}
where, when $f$ and $f'$ stand for quarks the $V_{L/R}$ represents the CKM 
matrix for the left and right-handed quarks respectively, whereas for the case of $f,f' \in$ leptons 
the $V_{L/R}$ represents the PMNS mixing matrix for the left-handed leptons and the corresponding 
analog for the right-handed leptons respectively. This interaction Lagrangian leads to the 
production of a heavy right-handed neutrino $N_R$ through the exchange of a 
right-handed charged gauge boson $W_R$ which constitutes the only dominant contribution 
to the process $$e^-p \to N_R ~j$$ at LHeC, where $j \in q_i$ (light quarks). Thus an $e^-p$ collider presents us with an 
opportunity of testing the LR symmetry breaking scale because of the direct coupling $e-N_R-W_R$
unique to the left-right symmetric theory.  
\begin{figure}[t!]
\includegraphics[width=2.9in,height=1.9in]{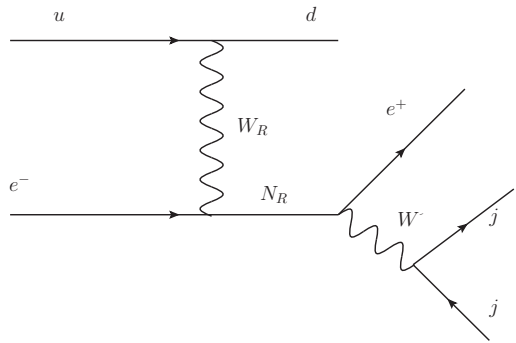}
\caption{Illustrating the Feynman diagram that contributes to the $N_{R} ~j + X$ production through the 
exchange of a heavy $W_R$ in the $t$-channel in $e^-p$ collision at LHeC.}
\label{feynfig}
\end{figure}

\emph{LHeC as a tool to confirm LR symmetry}:
As pointed out earlier, the search for a heavy right-handed neutrino ($N_R$) at the LHC through the  
LNV $\Delta L=2$ process is quite well studied. Unlike typical TeV-scale seesaw scenarios 
where the production is mediated by the SM $W$ boson the production in the LR model is mediated 
by the heavy $W_R$ gauge boson. We propose in this Letter that the LHeC would be an excellent place to 
determine such information and confirm the LR symmetry. On the downside, the LHeC loses the advantage 
of producing a heavy $W_R$ as a resonance like the LHC. Thus it might seem that it would be limited in 
the mass reach of $W_R$.  However an important thing to note here is that for an $ep$ machine such as 
the LHeC, an exchange of the right-handed $W_R$ leads directly to the production of a right-handed 
neutrino of the electron-type, which is only possible in the LR model (see Eq.~\ref{eq:lagrangian}). 
This invariably makes it the minimal process possible with a $W_R$ exchange to probe the LR scale.  
In terms of the production rates for $N_R$, the LHeC should definitely 
outperform future $e^+e^-$ colliders. Note that the production of a right-handed $N_R$ at LHeC also 
presents us with a process with negligible SM background\footnote{There may be some
non-trivial SM background induced by the jet-electron misidentification.} which can more than compensate for the 
suppressed rates compared to LHC. 
The process in consideration is shown in Fig.\ref{feynfig} where we get $$e^-p \to e^+~j +X$$ which 
is the LNV $\Delta L=2$ analog of what is expected at the LHC through a Majorana neutrino production. 
This mode for a heavy Majorana neutrino signal has been considered before for 
\begin{figure}[t]
\includegraphics[width=3.4in,height=2.8in]{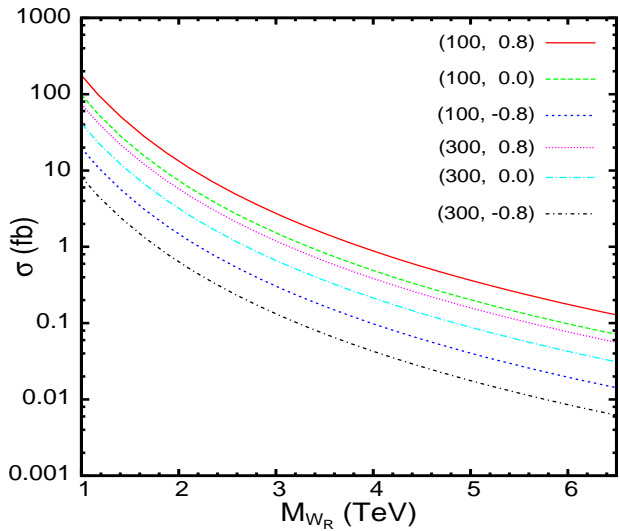}
\caption{The production cross section for the $N_R~j$ at LHeC as a function of the $W_R$ mass. Different 
curves correspond to different values of $N_R$ mass and the choice for electron beam polarizations 
represented as ($M_{N_R}, P_e$) in parentheses. }
\label{fig:cross}
\end{figure}
LHeC \cite{Liang:2010gm, Blaksley:2011ey, Duarte:2014zea}. We wish to highlight the fact that as the 
electron beam at the LHeC can be longitudinally polarized to study electroweak physics and to disentangle 
vector and axial vector contributions \cite{AbelleiraFernandez:2012cc}, this can be used to the advantage 
of studying the $N_R$ production which shall invariably also confirm any hint of a LR symmetry in nature. 

We note that the LHeC will be a ``ring-ring (RR)'' collider with an option for the 
``linac-ring (LiR)'' configuration too, with a linear electron accelerator tangential to the LHC.  The RR 
option will be able to provide high luminosity but for higher electron energies in the ring, 
maintaining the beam polarization is difficult. Therefore the beam polarization shall decrease as the 
electron beam energy is ramped up.  With a modest beam energy of 60 GeV for the electron beam one can 
still have 80\% L/R beam polarization, while the proton beam carries 7 TeV energy. 
The LiR configuration however has a high potential for polarised electrons.  We use the RR configuration 
to present our results noting that improved energies at LiR for the electron beam while maintaining the 
polarization would only help to enhance the signal rates further.

\begin{figure}[t]
\includegraphics[width=3.4in,height=2.8in]{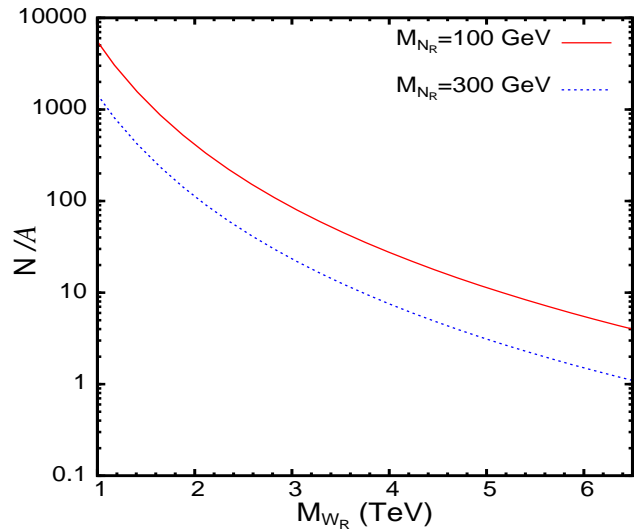}
\caption{The event rate at 100$fb^{-1}$ luminosity normalized with acceptance efficiency ($\mathcal A$) at LHeC as a function of the $W_R$ mass with 
the electron beam polarization  $P_e=0.8$. }
\label{fig:events}
\end{figure}

\emph{Analysis and results}: The $e^-p \to N_R~j$ production cross section is calculated in the LR symmetry model using the package 
{\tt CalcHEP} \cite{Belyaev:2012qa}. We have used the model files for the minimal model provided 
in Ref \cite{Roitgrund:2014zka}. For the initial quarks we use the {\tt CTEQ6L1} parton distribution functions 
(PDF) \cite{Pumplin:2002vw}. The proton has $E_p= 7$ TeV while the electron beam has $E_e = 60$ GeV. 
We take the maximum available electron beam polarization ($P_e$) to be 80\%. In Fig. \ref{fig:cross} we plot the 
production cross section of $e^-p \to N_R~j$ as a function of the $W_R$ mass for two different values of the 
heavy neutrino mass $M_{N_{R}}=100, 300$ GeV. We consider three different initial beam configurations for 
the electron, {\it i.e.} when it is left-polarized ($P_e=-0.8$), unpolarized  ($P_e= 0$) and 
right-polarized ($P_e = 0.8$). While estimating the production cross section, we set some basic acceptance 
cuts for the massless quarks representing $j$ produced at the parton level, 
as $p_T^j > 10$ GeV and $|\eta_j| < 4$. 
\begin{figure}[t!]
\includegraphics[width=3.4in,height=2.8in]{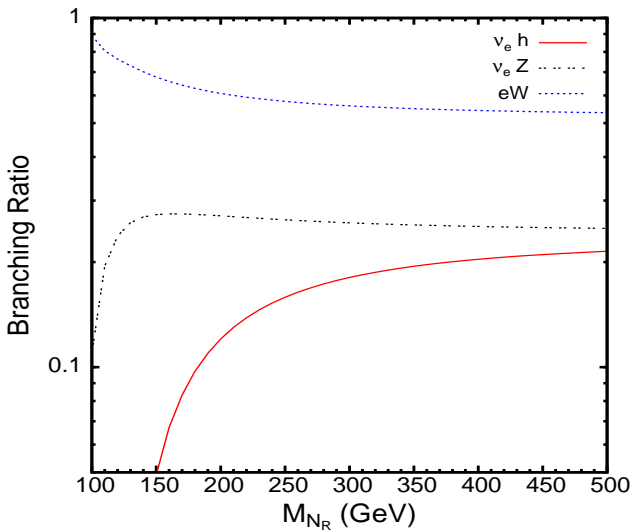}
\caption{Illustrating the relevant branching fractions of the right-handed neutrino $N_R$ decay modes
as a function of its mass. Note that the $eW$ mode represents the sum of equal contributions from 
$N_R \to e^+W^-$ and $N_R \to e^-W^+$. }
\label{fig:decay}
\end{figure}
As one would expect, due to the right-handed nature of interaction there is a clear enhancement in the 
production rates with the electron beam right-polarized, when compared to the unpolarized beam. 
In fact we find that the production rates for $N_R~j$ using an unpolarized electron beam with 
$E_e=110$ GeV is similar to the cross section for $E_e=60$ GeV and $P_e = 0.8$. In addition, the cross 
section would drop considerably once the beam polarization is reversed. Ideally for a 100\% left polarized 
beam there would hardly be any contribution to the $N_R$ production in the LR symmetric model. 
However, with 80\% left-polarized electron beam one is still able to get a small fraction of the $W_R$ 
contribution as shown in Fig. \ref{fig:cross}. We also find that LHeC could be sensitive to similar or even 
greater values of $W_R$ mass that could be excluded at the LHC, provided the right-handed neutrino is 
not too heavy.  For example, when $M_{W_R} = 6$ TeV and the electron beam is right-polarized 
($P_e=0.8$), we get $\sigma (N_R~j) \sim 0.08$ fb for $M_{N_R}=300$ 
GeV while  $\sigma (N_R~j) \sim 0.18$ fb for $M_{N_R}=100$ GeV. That corresponds to about 8 and 18 
events respectively, with an 100 $fb^{-1}$ integrated luminosity. We consider the $e^+ ~X $ final 
state which arises when 
$$ N_R \to e^+ W^- \to e^+ ~ jj,$$
where $W$ decays hadronically with $\sim 0.7$ branching fraction. A quick look at the decay modes of the 
heavy neutrino, shown in Fig. \ref{fig:decay} gives about 30\% branching for  $M_{N_R}=100$ GeV
and about 21\% for  $M_{N_R}=300$ GeV to $e^+ W^- \to e^+ 2j$ mode. This would give us nearly 
$5 \mathcal A$ events for $M_{N_R}=100$ GeV and $2 \mathcal A$ events for $M_{N_R}=300$ GeV 
respectively. Here we denote $\mathcal A$ as the acceptance efficiency for events after basic trigger cuts 
for the final state $e^+ X$, which we expect would be sufficiently high while putting basic acceptance cuts
for a final state with practically no SM background. 
The event rates are naturally greater as we go lower in the $W_R$ mass and we find that it nearly doubles 
when $M_{W_R}=5$ TeV. Note that the change in $M_{W_R}$ does not affect $N_R$ decay branchings much.  We show the event rates scaled by acceptance efficiency ($\mathcal A$) for the above final state 
in Fig. \ref{fig:events} for different $W_R$ mass.
Note that as there is {\it absolutely no SM background} for the above final state, an observation of 
even a very few distinct events would be a clear signal for the existence of a left-right
symmetry in nature. In addition a clear and complementary confirmation would come through the vanishing 
of the same signal events once the electron beam polarization is flipped. Note that in other scenarios of heavy 
Majorana neutrino with no right-handed symmetry, the couplings are left-handed in nature and therefore a 
right-polarized beam would give no signal events while a left-polarized beam will enhance such a signal.  
If high enough beam polarization could be maintained for higher electron energies too, that would clearly help 
in increasing the rates further. 
  
\emph{Summary and Conclusions:}
To summarize, in this Letter we have shown that the proposed $ep$ collider, LHeC could be an 
excellent place to test the existence of LR symmetry in nature and understand why parity symmetry is 
maximally violated in weak interactions. LR model is the most natural extension to the SM where 
right-handed neutrinos appear naturally and also help us understand that parity as a symmetry 
is broken at a high scale. A  TeV scale breaking leads to interesting collider signals.  Although much 
lower energies compared to the LHC would be available, we show that the sensitivity can be quite high 
and comparable with the LHC through the production of a right-handed heavy neutrino. We also find that using a longitudinally polarized 
electron beam will be crucial in the search for LR symmetry at LHeC. The correct polarization not only 
enhances the right-handed signal but also distinguishes it from any competing scenarios of neutrino mass 
generation models, which is difficult at the LHC. There could also be other possible new physics 
studies, using the polarized electron beam option at LHeC  \cite{Cakir:2014swa}.
We must however acknowledge the fact that the LR signal might not show once the right handed 
neutrinos also become too heavy. In such a scenario, to maintain a polarized beam with higher energies, 
one would have to use the linac type setup (LiR) for the electron beam \cite{Kaya:2015tia}.  
For a linac solution the electron beam polarization would be independent of beam energy and therefore 
a polarized electron beam with higher energies can be used at such $ep$ colliders. 
Thus we find that having the electron beam polarized at LHeC not only helps in studying the electroweak 
physics within SM, but can prove to be a novel feature in unravelling one of the fundamental mysteries {i.e.} 
Parity as a symmetry in nature and the existence of left-right symmetry.   

\begin{acknowledgments}
\emph {Acknowledgments:} 
This work was partially supported by funding available from the Department of Atomic Energy, Government 
of India, for the Regional Centre for Accelerator-based Particle Physics (RECAPP), Harish-Chandra Research 
Institute.
\end{acknowledgments}


\end{document}